\documentclass[orivec]{llncs}

\usepackage{units} 

\usepackage{tikz}
\usetikzlibrary{matrix,arrows}
\tikzstyle{every picture}+=[descr/.style={fill=white,inner sep=2.5pt},node distance=5.5em]
\newcommand{\obj}[3]{\node (#1) [#2] {$#3$};}
\newcommand{\arr}[3]{\path[->,font=\scriptsize](#2) edge node[auto] {$#1$} (#3);}

\usepackage{amsmath}
\usepackage{amsfonts}
\usepackage{amssymb}

\usepackage{verbatim}

\newcommand{\mC}[1]{\mathtt{#1}}
\newcommand{\mF}[1]{\mathtt{#1}}
\newcommand{\Set}{\mC{Set}}
\newcommand{\Id}{\mF{Id}}

\newcommand{\E}{\mC{E}}
\newcommand{\F}{\mF{F}}
\newcommand{\G}{\mF{G}}
\newcommand{\B}{\mF{B}}
\newcommand{\U}{\mF{U}}
\newcommand{\T}{\mF{T}}
\newcommand{\K}{\mF{K}}
\newcommand{\reach}{\mC{R}}

\newcommand{\id}{\mathit{id}}
\newcommand{\cod}{\mF{cod}}
\newcommand{\dom}{\mF{dom}}

\newcommand{\dialg}{\mathit{Dialg}}
\newcommand{\coalg}{\mathit{Coalg}}

\newcommand{\tr}[1]{\stackrel{#1}{\longrightarrow}}
\newcommand{\utr}{\to}

\newcommand{\Pow}{\mathcal{P}}

\newcommand{\R}{\mathcal{R}}
\newcommand{\sbb}{\stackrel{\ldotp}{\sim}}
\newcommand{\sbe}{\simeq}

\newcommand{\fn}{\mathit{fn}}

\newcommand{\chan}{\mathcal{C}}

\newcommand{\defend}{}
\newcommand{\proend}{\medskip}

\renewenvironment{definition}{\begin{defn}}{\defend\end{defn}}

\renewenvironment{theorem}{\begin{thm}}{\defend\end{thm}}

\renewenvironment{proof}{\begin{pro}}{\proend\end{pro}}

\renewenvironment{corollary}{\begin{cor}}{\defend\end{cor}}

\usepackage{dashrule}

\begin{document}

\title{Interaction and observation: categorical semantics of reactive systems trough dialgebras}
\author{Vincenzo Ciancia\thanks{The research leading to these results was partially supported by
the EU FET 7FP Collaborative Project n. 600708 (QUANTICOL), the PAR FAS 2007-2013 (TRACE-IT) project, and the EU 7FP
projects n. 256980 (NESSoS), n. 257930 (Aniketos), n. 295354 (SESAMO).} }
\institute{Istituto di Scienza e Tecnologie dell'Informazione ``A. Faedo''\\ 
Consiglio Nazionale delle Ricerche \\ Pisa, Italy}
\maketitle
\begin{abstract}
We use dialgebras, generalising both algebras and coalgebras, as a complement of the standard coalgebraic framework, aimed at describing the semantics of an interactive system by the means of reaction rules. In this model, interaction is built-in, and semantic equivalence arises from it, instead of being determined by a (possibly difficult) understanding of the side effects of a component in isolation. Behavioural equivalence in dialgebras is determined by how a given process interacts with the others, and the obtained observations. We develop a technique to inter-define categories of dialgebras of different functors, that in particular permits us to compare a standard coalgebraic semantics and its dialgebraic counterpart. We exemplify the framework using the CCS and the $\pi$-calculus. Remarkably, the dialgebra giving semantics to the $\pi$-calculus does not require the use of presheaf categories.
\end{abstract}

\section{Introduction}

A system is called \emph{interactive} when its semantics depends upon interaction with a surrounding environment. The semantics does not just yield a value (or not at all), but rather it consists in the denotation of the \emph{behaviour} of the system itself, usually described either by \emph{reaction rules} or by a \emph{labelled transition system} (LTS). The difference is illustrated by the following example, defining a reaction rule for the synchronisation of two parallel processes in a process calculus (the rule on the left) or the LTS  variant (the three rules on the right):
$$ {a . P \parallel \bar a . Q \utr P \parallel Q} \quad \quad \quad {a . P \tr a P} \quad {\bar a . P \tr {\bar a} P} \quad \frac{P \tr{a} P' \,\,\, Q \tr{\bar a} Q'}{P \parallel Q \tr{\tau} P' \parallel Q'}$$
Here $a . P$ is a process waiting for a signal on channel $a$, whose continuation is $P$. Similarly, $\bar a .P$ sends a signal on $a$, while $P \parallel Q$ is the parallel composition of two processes. The reaction rule may be read as ``whenever two processes can synchronise, they do, and evolve into the parallel composition of their continuations''. The LTS rules may be read as: ``whenever a process can send or receive a signal, it evolves into its continuation, and has a \emph{side effect} on the environment''. Two processes with complementing side effects interact by the last rule. 


LTSs are widely used for modelling the semantics of interactive systems, since they come equipped with bisimilarity, a form of \emph{behavioural equivalence} that specifies when the semantics of two processes is the same. Coalgebras generalise LTSs and have a standard definition of bisimilarity, coinciding with kernel equivalence of morphisms under mild assumptions. Many formalisms received a coalgebraic treatment. This becomes increasingly harder as the complexity of the calculus grows, depending on the general question of \emph{what are side effects} in a specific calculus. For example, dealing with name allocation in the $\pi$-calculus requires the use of presheaf categories (see e.g. \cite{ft99}). 
%

%
%
In this work, we seek for a setting where reaction rules are the main object of study, and side effects or labels are not needed at all to define behavioural equivalence.
We
use a generalisation of coalgebras, namely \emph{dialgebras}, to represent a rule system as an object in a category, so that kernel equivalence can be used as a notion of behavioural equivalence.
%
%
%
%
To appreciate the difference, consider a function $f : X \to \Pow(L \times X)$. It specifies an LTS with states in $X$ describing, for each state $x$, the non-deterministic choices at $x$, the side effects of each choice, and the resulting state. In LTSs, and coalgebras, elements are observed in isolation. In contrast, an example of a dialgebra is $f : X \times X \to \Pow(X)$. The value of $f(x,y)$ is meant to describe the possible (non-deterministic) outcomes of an interaction between $x$ and $y$. Elements are not observed in isolation, but rather their mutual interactions define the semantics.

Categories of dialgebras were first studied under the name of \emph{generalised algebraic categories} (see e.g. \cite{TG69,Ada76}). These structures have been used in computer science for the specification of data types \cite{Hag87,PZ01}. A systematic study of dialgebras, patterned after \cite{Rut00}, was done in \cite{Vou10}. Applications to a compositional calculus for software components were proposed in \cite{CiE11}.  In \cite{Cia11}, we modelled asynchronous process calculi as isolated machines that can be fed with input tokens by an external observer, using dialgebras to generalise \emph{Mealy machines}. In this paper we aim at a more intensional characterisation, tailored to reaction rules, obtained by studying interaction between pairs of systems rather than the relation between their input and output. Our work diverts from previous research, as we take a ``local'' approach. Instead of studying a whole category of dialgebras, 
one just considers objects that are reachable by morphisms from the particular dialgebra being studied.
This approach addresses the issue, previously unsolved, that a final dialgebra fails to exist, so there is no universal semantic domain for the whole class of considered objects. To obviate to this, in \S \ref{sec:dialgebras}, we study the \emph{bisimilarity quotient} of a specific system, showing that it exists under mild conditions. The bisimilarity quotient is sufficient to provide a canonical semantics to an interactive system up-to behavioural equivalence, by the means of a canonical epimorphism. Indeed, a universal model is also useful to compare different systems of the same type. However, this is not typical in coalgebraic program semantics, as different systems have different types, representing different kinds of side effects. In dialgebraic program semantics, just like in the coalgebraic case, working with different systems requires the use of categorical comparisons. In \S \ref{sec:comparing-all}, we develop techniques to do so in the local perspective.

In \S \ref{sec:examples} we provide two examples: the \emph{Calculus of Communicating Systems} and the $\pi$-calculus. The simplicity of the former allows us to cleanly illustrate our framework, including the comparisons of \S\ref{sec:comparing-all}, used to prove the equivalence of the coalgebraic and dialgebraic semantics. On the other hand, the dialgebraic semantics of the $\pi$-calculus provides an important insight: using reaction rules, one does not need to understand what are the side effects of programs, as interaction is sufficient to characterise their behaviour. Thus, the semantics of the calculus is given as a dialgebra in $\Set$, without the need to resort to presheaf categories to describe bisimilarity; but nevertheless, we are able to prove that standard dialgebraic bisimilarity coincides with early bisimilarity. It is also worth notice that the close relationship between the two calculi permits us to define the dialgebra for the $\pi$-calculus by just changing one rule from the dialgebra for the CCS.  This is possible since both systems can be described by the means of binary \emph{interactions} and non-deterministic \emph{observations}; that is, they use the same functors. We introduce these two simple functors in \S \ref{sec:interaction-and-observation-functors}, together with a characterisation theorem for the associated behavioural equivalence.


\paragraph{Related work.} The most widely known framework for the categorical semantics of reactive systems is the so-called ``\emph{contexts as labels}'' approach \cite{lm00,Sew02,ss03}. Roughly, an LTS is derived from reaction rules adopting unary contexts as labels; bisimilarity of such LTS serves as behavioural equivalence.
Nevertheless, \emph{minimal} contexts need to be carefully selected in order to obtain a sensible equivalence relation; the obtained semantics depends on this choice, and is not directly specified by the reaction rules themselves. 
The resulting categorical framework is highly non-trivial.
%
Dialgebras, in contrast, provide a simple setting in which to study reactive systems. However, compositionality, which is a foundational reason to use contexts as labels, and certainly a desired feature, has not yet been investigated for dialgebras (more on this in the conclusions). For the time being, dialgebras complement algebras and coalgebras, rather than \emph{bialgebras}. 

\section{Dialgebras}
\label{sec:dialgebras}

A coalgebraic semantics can be interpreted as the discerning power of an \emph{observer} that can see all the actions done by a process. In contrast,  \emph{dialgebras} endow the observer with the ability to \emph{interact} with the system. The passive observer becomes an entity which runs \emph{experiments} and \emph{observes} the results. By this change of point of view, we can represent e.g. input as an experiment in which the observer feeds a system with a value \cite{Cia11}, or we can represent interaction as a binary experiment involving two processes as we do in the current paper.

We restrict all our definitions to the category $\Set$, but indeed the theory of dialgebras can be developed in any category.

\begin{definition}\label{def:dialgebra} Given two functors $\F, \B : \Set \to \Set$, a $(\F,\B)$-\emph{dialgebra} is a pair $(X,f)$ where $X$ is the \emph{carrier set} or \emph{underlying set} and $f : \F X \to \B X$ is a function. A $(\F,\B)$-dialgebra homomorphism from $(X,f)$ to $(Y,g)$ is a function $h : X \to Y$ such that $g \circ \F h = \B h \circ f$, as depicted in Figure \ref{fig:dialgebra-homomorphism}. Dialgebras and their homomorphisms form the category $\dialg(\F,\B)$.
\end{definition}

\begin{figure}
	\begin{center}
		\begin{tikzpicture}
			\obj{X}{}{X}
			\obj{Y}{below of = X}{Y}
			
			\arr{h}{X}{Y}
			 
			\obj{FX}{right of = X}{\F X}
			\obj{BX}{right of = FX}{\B X}
			\obj{FY}{below of = FX}{\F Y}
			\obj{BY}{right of = FY}{\B Y}
			
			\arr{\F h}{FX}{FY}
			\arr{\B h}{BX}{BY}
			\arr{f}{FX}{BX}
			\arr{g}{FY}{BY}
	    \end{tikzpicture}
	\end{center}
	\caption{A dialgebra homomorphism.} \label{fig:dialgebra-homomorphism}
\end{figure}


Roughly, $\F$ is the syntax of experiments, of which the function $f$ is the semantics, yielding a set of elements in a type of observed results $\B$. The crucial feature of dialgebras is that, since they form a category, they have a standard notion of equivalence coming from kernels of morphisms. Notice that dialgebras conservatively extend algebras ($\dialg(\F,\Id)$) and coalgebras ($\dialg(\Id,\T)$).

\begin{definition}\label{def:dialgebraic-bisimilarity}
	Given a $(\F,\B)$-dialgebra $(X,f)$, \emph{dialgebraic bisimilarity} $\sim_f \subseteq X \times X$ is defined by $x \sim_f y \iff \exists (Y,g) . \exists h : (X,f) \to (Y,g) . h(x) = h(y)$. 
\end{definition}
\medskip

Notice that in Definition \ref{def:dialgebraic-bisimilarity} we use the kernel of $h$ as a function. Thus, we do not require kernels (pullbacks of a function with itself) in $\dialg(\F,\B)$. The ``extensional'' definition that we provide is applicable to any kind of dialgebra (independently from $\F$ and $\B$), and it avoids the machinery of relation liftings (which are used in \cite{PZ01}). We now study epi-mono factorisations of dialgebras. 

\begin{proposition}\label{prop:epi-mono-factorisations}
	If $\F$ and $\B$ preserve monos whose domain is empty, the category $\dialg(\F,\B)$ has unique epi-mono factorisations. Otherwise, it has epi-mono factorisations of morphisms whose domain does not have an empty carrier.
\end{proposition}

Proposition \ref{prop:epi-mono-factorisations} guarantees that bisimilarity is determined by the  epimorphisms.  
%
In coalgebras, the kernel of the unique morphism into the final object (if any) coincides with bisimilarity. For simple functors $\F$, e.g., $\F(X) = X\times X$, even when $\B$ is bounded, a final dialgebra does not exist\footnote{A final dialgebra still exists when $\B$ preserves the terminal object \cite{Vou10}. However, this makes the category of dialgebras not very interesting, as the final dialgebra has just one element, thus all the elements of any system are bisimilar.}.

\begin{example}\label{exa:lack-of-final-dialgebra}
%
%
Let $\F(X) = X \times X$ and $\B(X) = \Pow_{fin}(X)$ (the finite power set of $X$). Suppose there is a final dialgebra $(Z,z)$. Consider the dialgebra $(Z+1,f)$ where $f(x,y) = z(x,y)$ if $x,y \in Z$, $f(x,*) = f(*,x) = \{ * \}$ if $x \in Z$, and $f(*,*) = \emptyset$. Consider the final map $h : (Z+1,f) \to (Z,f)$. Here we get to a contradiction: $h$ restricted to $Z$ must be the identity, therefore injective and surjective. To see this, consider that $(Z,z)$ embeds into $(Z+1,f)$ by the identity function, and that the identity of $(Z,z)$ must factor through such embedding and $h$, by finality. Thus, there is $x \in Z$ such that $h(x) = h(*)$. Then $h$ is not a dialgebra homomorphism: we have $z(\F h(*,x)) = z(h(*),h(x)) = z(h(*),h(*)) = z(\F h(*,*))$, while $\B h (f(*,x)) = \{ h(*) \} \neq \emptyset = \B h(f(*,*))$.
Intuitively, the element $*$ behaves differently from every other element of $Z$, but $Z$ ought to encompass all the possible behaviours, which is a contradiction.
Similarly, for any $X$, define the dialgebra $g(x,x) = \{x\}$, $g(x,y) = \emptyset$ if $x \neq y$. Since $X$ is arbitrary, and no different elements are bisimilar, the cardinality of a final dialgebra is unbounded.
\end{example}

Final semantics is a well-established way to define behavioural equivalence of systems. 
In the absence of a final object in $\dialg(\F,\B)$, we can still define behavioural equivalence by reasoning in terms of quotients of a system. Recall that a \emph{quotient} of an object $X$ in a category is the canonical representative of an equivalence class of epimorphisms from $X$, under the equivalence relation $f : X \to Y \equiv g : X \to Z$ if and only if there is an isomorphism $i : Y \to Z$ such that $i \circ f = g$. In $\Set$, the quotients of an object form a set. 

\begin{definition}
 The \emph{bisimilarity quotient} of a dialgebra $(X,f)$ is the wide pushout $(Q,q)$ (if it exists) of the cone of \emph{quotients} of $(X,f)$ in $\dialg(\F,\B)$. We call the diagonal $z : (X,f) \to (Q,q)$ the \emph{canonical map} of $(X,f)$.
\end{definition}

\begin{proposition}\label{pro:bisimilarity-equivalence}
	Let $(Q,q)$ be the bisimilarity quotient of $(X,f)$ and $z$ the canonical map. For all $x,y \in X$, $x$ is bisimilar to $y$ if and only if $z(x) = z(y)$. Therefore, when the bisimilarity quotient exists, bisimilarity is an equivalence relation.
\end{proposition}

When the bisimilarity quotient exists, the canonical map can be considered the semantics of a system. Although $z$ is canonical, it is not necessarily the unique $z : (X,f) \to (Q,q)$: bisimilarity classes may be interchangeable in dialgebras. 

\begin{example}Consider the dialgebra $g$ defined in Example \ref{exa:lack-of-final-dialgebra}; the dialgebra has no non-trivial quotients, so it coincides with its bisimilarity quotient. However all the isomorphisms of the carrier set are dialgebra homomorphisms. The bisimilarity classes are the same, but for the identity of the sole element of each class. 
\end{example}

For a different example, in the semantics of a symmetric process calculus such as the \emph{pure} variant of CCS, the bisimilarity classes of an element, and of the element obtained by replacing all the input or output actions in it with the complementary ones, can be interchanged.
%

Clearly, when a final object exists, the epi-mono factorisation of the final morphism yields the bisimilarity quotient. A bisimilarity quotient may exist also in the absence of a final object. Here is a sufficient condition.


\begin{proposition}\label{pro:existence-bisimilarity-quotients}
	When $\F$ preserves wide (small) pushouts of epimorphisms, that is, colimits of an arbitrary small cone of epis, the bisimilarity quotient exists.
\end{proposition}

The dialgebraic semantics of CCS in \S \ref{sec:CCS-dialgebra} is an example where the bisimilarity quotient exists, but there is no final dialgebra (by Example \ref{exa:lack-of-final-dialgebra}).

\section{Interaction and observation}
\label{sec:interaction-and-observation-functors}

In this section we introduce two functors $\F$ and $\B$ that can be used to give dialgebraic semantics to interactive, non-deterministic calculi.

\begin{definition}\label{def:int-and-obs-functors}
	The \emph{interaction} and \emph{observation} functors are defined as $\F(X) = X + (X \times X)$ and $\B(X) = \Pow(X)$, respectively.
\end{definition}

As elements of $X$ and $X \times X$ are syntactically disjoint, we use them to denote elements of $\F(X)$, avoiding labels for the coproduct. An element of $\F X$ is either $x \in X$, representing an experiment about a process in isolation, or $(x,y) \in X \times X$, an experiment where two processes are allowed to interact. Elements of $\B$ are sets of processes, that are the possible non-deterministic outcomes of an experiment. We write $x \utr z$ and $(x,y) \utr z$ for $z \in f(x)$ and $z \in f(x,y)$, respectively.

\begin{proposition}\label{pro:int-obs-epi-mono-bisim-quotient}
	The functors $\F$ and $\B$ preserve monos from the empty set. $\F$ preserves wide pushouts of epis. Therefore $\dialg(\F,\B)$ has epi-mono factorisations, and each dialgebra in the category has a bisimilarity quotient.
\end{proposition}

\noindent Theorem \ref{thm:bisimilarity-back-and-forth} provides a characterisation of bisimilarity in $\dialg(\F,\B)$.

\begin{theorem}\label{thm:bisimilarity-back-and-forth}
 Let $(X,f)$ be an $(\F,\B)$-dialgebra. An equivalence relation $\R\subseteq X \times X$ is the kernel of $h : (X,f) \to (Y,g)$ for some $h$, $Y$, $g$, if and only if, for all $(x_1,x_2), (y_1,y_2) \in \R$ and $z_1\in X$, we have: $x_1 \utr z_1 \implies \exists z_2 . x_2 \utr z_2 \land (z_1,z_2) \in \R$, and, $(x_1,y_1) \utr z_1 \implies \exists z_2 . (x_2,y_2) \utr z_2 \land (z_1, z_2) \in \R$. As a corollary, the bisimilarity quotient of $(X,f)$ is the largest such relation.
\end{theorem}

%
%
%

\section{Comparing dialgebras}\label{sec:comparing-all}


Categories of algebras or coalgebras of different functors may be compared by mapping one category into the other by composition with an appropriate natural transformation \cite{Rut00}. In \S \ref{sec:comparing-categories} we generalise this technique to dialgebras (of which algebras and coalgebras are special cases). The problem has first been studied in \cite{Vou10}. Here we improve on it by adding an intermediate ``container'' functor $\G$, whose role is crucial, e.g., for \S \ref{sec:ccs-comparison}. In \S \ref{sec:specific-comparisons} we discuss the limitations of this method in the setting of dialgebras, and refine the construction.

\subsection{Comparing categories of dialgebras}
\label{sec:comparing-categories}

\begin{figure}
\begin{center}
\begin{tikzpicture}
	\obj{F1X}{}{\F'X}
	\obj{F1Y}{below of = F1X}{\F'Y}
	\obj{GFX}{right of = F1X}{\G \F X}
	\obj{GFY}{right of = F1Y}{\G \F Y}
	\obj{GBX}{right of = GFX}{\G \B X}
	\obj{GBY}{right of = GFY}{\G \B Y}
	\obj{B1X}{right of = GBX}{\B'X}
	\obj{B1Y}{right of = GBY}{\B'Y}

	\arr{\F' h}{F1X}{F1Y}
	\arr{\G \F h}{GFX}{GFY}
	\arr{\G \B h}{GBX}{GBY}
	\arr{\B' h}{B1X}{B1Y}
	
	\arr{\lambda_X}{F1X}{GFX}
	\arr{\lambda_Y}{F1Y}{GFY}
	\arr{\mu_X}{GBX}{B1X}
	\arr{\mu_Y}{GBY}{B1Y}
	\arr{\G f}{GFX}{GBX}
	\arr{\G g}{GFY}{GBY}
	
	\obj{FX}{right of = B1X}{\F X}
	\obj{FY}{below of = FX}{\F Y}
	\obj{BX}{right of = FX}{\B X}
	\obj{BY}{right of = FY}{\B Y}
	
	\arr{\F h}{FX}{FY}
	\arr{\B h}{BX}{BY}
	\arr{f}{FX}{BX}
	\arr{g}{FY}{BY}
	
	\obj{X}{left of = F1X}{X}
	\obj{Y}{below of = X}{Y}
	
	\arr{h}{X}{Y}
\end{tikzpicture}	
\end{center}
\caption{Comparing categories of dialgebras using natural transformations.}\label{fig:natural-in-set}
\end{figure}
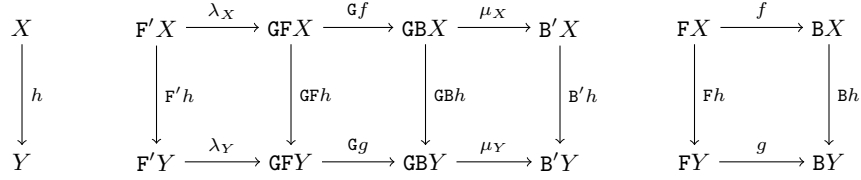

Consider $\Set$ endofunctors $\F,\B,\F',\B'$. One can specify functor $\K : \dialg(\F,\B) \to \dialg(\F',\B')$ using $\G : \Set \to \Set$ and two natural transformations $\lambda : \F' \to \G \F$ and $\mu : \G \B \to \B'$, as illustrated by the diagram in Figure \ref{fig:natural-in-set}.

Notice that dialgebras come equipped with the ``underlying set'' or ``forgetful'' functor $\U_{\F,\B} : \dialg(\F,\B) \to \Set$ defined as $\U_{\F,\B}(X,f) = X$, $\U_{\F,\B}(h : (X,f) \to (Y,g)) = h$; this allows us to state that $\K$ is \emph{concrete}.

\begin{theorem}{\label{thm:general-comparisons}}
	Two natural transformations $\lambda : \F' \to \G \F$ and $\mu : \G \B \to \B'$ determine a functor $\K: \dialg(\F,\B)\to \dialg(\F',\B')$ as $\K(X,f) = (X,\mu_X \circ \G f \circ \lambda_X)$, $\K(h : (X,f) \to (Y,g)) = h$. $\K$ is concrete, that is: $\U_{\F',\B'} \circ \K = \U_{\F,\B}$. 
\end{theorem}	
	
As a consequence of $\K$ being concrete, we have the following corollary.

\begin{corollary}
	If $x,y\in X$ are bisimilar in $(X,f)$ then they are so in $\K(X,f)$.
\end{corollary}

\subsection{Comparing dialgebras}
\label{sec:specific-comparisons}

The framework of \S \ref{sec:comparing-categories} is more restrictive than necessary. Observe that $\lambda$ and $\mu$ have to be defined for each set $X$. But when comparing say, two different semantics of a language, we are only interested in the two dialgebras, and in the objects that may be reached from them by an epimorphism. Considering this subclass of objects, in the definition of $\lambda$ and $\mu$, we could use specific features of a given dialgebra (e.g. exploit the fact that the underlying set $X$ is the carrier of an algebra, or refer to specific elements of $X$).
In this light, we now restrict our attention to natural transformations between functors whose codomain is $\Set$, but whose domain is not. The framework appears complicated at first; however, Theorem \ref{pro:nat-uniquely-determined} allows us to specify such natural transformations by single functions, called \emph{bisimulation invariants}, that serve as an ``adaptation layer'' between dialgebras of different type.
Although proofs in this section require quite a bit of categorical reasoning, defining a bisimulation invariant is not a difficult task by itself and encapsulates the complexity of the framework in a simple definition. We shall support this claim in \S \ref{sec:ccs-comparison}, which proves equivalence of the dialgebraic and coalgebraic semantics of CCS. 


Let $\E$ be the subcategory of epimorphisms of $\dialg(\F,\B)$. Given a dialgebra $(X,f)$, consider the coslice category $(X,f) / \E$. Its objects are arrows in $\E$ whose domain is $(X,f)$. Its arrows are commuting morphisms of $\E$. Our framework is parametrised by a full subcategory of $(X,f) / \E$ having the following properties.

\begin{definition}\label{def:reach}
	Let $\reach_{(X,f)}$ be a full subcategory of $(X,f) / \E$, such that:
		for each object $h$ of $(X,f) / \E$ there is at least one object $h'$ of $\reach_{(X,f)}$ with a commuting arrow $k : h \to h'$;
	    %
	    the identity $\id_{(X,f)}$ is an object.
\end{definition}

By the first condition, $\reach_{(X,f)}$ contains enough epimorphisms to characterise bisimilarity: whenever $x$ and $y$ are identified by a morphism $h$, there is $h'$ in the category that identifies them. The second condition embeds the object $(X,f)$ into $\reach_{(X,f)}$. The purpose of $\reach_{(X,f)}$ is to serve as a domain for functors into $\Set$, so that natural transformations between them may be more specific, while preserving bisimilarity of the object $(X,f)$. A natural transformation in this setting may make explicit reference to $(X,f)$. Notice that the maps composing such a natural transformation depend on dialgebra morphisms, not objects, by definition of $\reach_{(X,f)}$. Using a \emph{subcategory} of $(X,f) / \E$ further relaxes proof obligations; e.g., in Proposition \ref{prop:invariants-CCS} we only consider morphisms whose kernels are congruences with respect to the parallel operator. 
The following definition casts $\Set$ endofunctors into functors from $\reach_{(X,f)}$ to $\Set$.
\begin{definition}\label{def:lifting} For each $\F: \Set \to \Set$, define its \emph{lifting} $\bar \F : \reach_{(X,f)} \to \Set$ as $\bar \F = \F \circ \U_{\F,\B} \circ \cod$, where $\cod : (X,f) / \E \to \E$ is the codomain functor, mapping objects (arrows in $\dialg(\F,\B)$) to their codomains, and arrows to themselves.
\end{definition}

Spelling out the definition, $\bar F$ acts on objects as $\bar \F (p : (X,f) \to (Y,g)) = F(Y)$ and on arrows as $\bar F(k) = k$. 
%
%
%
%
Next, we prove that natural transformations indexed by $\reach_{(X,f)}$ may be specified by single functions, obeying to a condition that we call \emph{bisimulation invariance}. Notice that, since $\reach_{(X,f)}$ is a full subcategory containing the identity of a coslice category, each arrow $h : (X,f) \to (Y,g)$ can be regarded as both an object of $\reach_{(X,f)}$ and an arrow in the same category from $\id_{(X,f)}$ to $h$ itself. In the following, we refer to the arrow as $\hat h$ to avoid confusion.
\begin{definition}\label{def:invariant}
Given two functors $\F,\G : \reach_{(X,f)} \to \Set$, consider a function $k : \F (\id_{(X,f)}) \to \G (\id_{(X,f)})$. Call $k$ \emph{bisimulation invariant with respect to $(X,f)$ and $\reach_{(X,f)}$ from  $\F$ to $\G$} if and only if, for all  $x_1,x_2 \in X$, and for each arrow $\hat h : \id_{(X,f)} \to h$ in $\reach_{(X,f)}$, we have $\F \hat h (x_1) = \F \hat h (x_2) \implies \G \hat h ( k (x_1)) = \G \hat h (k (x_2))$. 
\end{definition}

In the following we call $k$ simply \emph{invariant} when $(X,f)$, $\reach_{(X,f)}$, $\F$ and $\G$ are clear from the context. Such a property of a function may seem difficult to prove. However, it is actually easier to show invariance of a given function with respect to the lifted versions of two functors, than naturality of a transformation in $\Set$ between the same functors. One needs to prove commutativity just for a given class of morphisms (those in $\reach_{(X,f)}$), which are also guaranteed to preserve and reflect bisimilarity. The fundamental property of invariants is Theorem \ref{pro:nat-uniquely-determined}, that depends on $\F$ preserving epis. This holds for the lifting $\bar \F$ of a $\Set$ endofunctor, as all $\Set$ endofunctors preserve epis, and so do $\cod$ and $\U$ used in Definition \ref{def:lifting}. 

\begin{theorem}\label{pro:nat-uniquely-determined}
	Consider two functors $\F,\G :  \reach_{(X,f)} \to \Set$, with $\F$ preserving epis. There is a one-to-one correspondence between natural transformations $\delta : \F \to \G$ and invariants from $\F$ to $\G$. Each natural transformation $\delta$ is uniquely determined by $\delta_{\id{(X,f)}}$, which is invariant; conversely, for each invariant $k$ there is a unique natural transformation $\delta$ such that $\delta_{\id{(X,f)}} = k$.
\end{theorem}

We can restate Theorem \ref{thm:general-comparisons} in terms of natural transformations $\lambda,\mu$ between functors from $\reach_{(X,f)}$ to $\Set$; this is described by the diagram in Figure \ref{fig:natural-in-reach-X-f}.
\begin{figure}
\begin{center}
	\begin{tikzpicture}[node distance=8em]		
		\obj{F1X}{}{\bar{\F'}(\id) = \F' X}
		\obj{F1Y}{below of = F1X}{\bar{\F'}(h) = \F' Y}
		\obj{GFX}{right of = F1X}{\G \bar{\F} (\id) = \G \F X}
		\obj{GFY}{below of = GFX}{\G \bar{\F} (h) = \G \F Y}
		\obj{GBX}{right of = GFX}{\G \bar{\B} (\id) = \G \B X}
		\obj{GBY}{below of = GBX}{\G \bar{\B} (h) = \G \B Y}
		\obj{B1X}{right of = GBX}{\bar{\B'}(\id) = \B' X}
		\obj{B1Y}{below of = B1X}{\bar{\B'}(h) = \B' Y}

		\arr{\bar{\F'} h = \F' h}{F1X}{F1Y}
		\arr{\G \bar{\F} h = \G \F h}{GFX}{GFY}
		\arr{\G \bar{\B} h = \G \B h}{GBX}{GBY}
		\arr{\bar{\B'} h = \B' h}{B1X}{B1Y}
		
		\arr{\lambda_{\id}}{F1X}{GFX}
		\arr{\lambda_h}{F1Y}{GFY}
		\arr{\mu_{\id}}{GBX}{B1X}
		\arr{\mu_h}{GBY}{B1Y}
		\arr{\G f}{GFX}{GBX}
		\arr{\G g}{GFY}{GBY}
		
	\end{tikzpicture}	
\end{center}
\caption{Comparing dialgebras using functors from $\reach_{(X,f)}$ to $\Set$. 	Here $\id$ is $\id_{(X,f)}$.} \label{fig:natural-in-reach-X-f}
\end{figure}	

\begin{theorem}\label{thm:specific-comparisons}
	Given an $(\F,\B)$-dialgebra $(X,f)$, a category $\reach_{(X,f)}$ as in Definition \ref{def:invariant}, a functor $\G$, and two invariants $\lambda$ from $\bar {\F'}$ to $\G \bar \F$, $\mu$ from $\G \bar \B$ to $\bar{\B'}$, consider the $(\F',\B')$-dialgebra $(X,f^{\lambda,\mu})$ where $f^{\lambda,\mu}=\mu \circ \G f \circ \lambda$. Whenever two elements are bisimilar in $(X,f)$, then they are bisimilar in $(X,f^{\lambda,\mu})$. 
\end{theorem}

\section{Examples}
\label{sec:examples}
\subsection{The coalgebraic semantics of CCS}
\label{sec:CCS-coalgebra}

The \emph{Calculus of Communicating Systems} (CCS) is a simple process calculus, formalising a fundamental aspect of computation: \emph{communication} between parallel processes. In the \emph{pure} variant only \emph{synchronisation} is considered, that is, the exchanged data is not taken into account.
We briefly recall the LTS (thus, coalgebraic) semantics of CCS here. The interested reader may refer to \cite{Mil82} for more details.
The syntax is described by the grammar: 
$$ P ::= \sum_{i \in I} \alpha_i . P_i \mid P_1 \parallel P_2 \mid (\nu a) P \qquad \qquad \alpha ::= \tau \mid a \mid \bar a$$
where $I$ is a finite set, and $a$ ranges over a countable set of channels $\chan$. Elements of $P$ are \emph{processes}, or \emph{agents}. Elements of $\alpha$ are \emph{atomic actions}, or \emph{prefixes}, or \emph{guards}.  CCS features operators for denoting: \emph{parallel composition} ($P_1 \parallel P_2$); \emph{restriction} of a channel $x$ which becomes private to $P$ ($(\nu a) P$);  non-deterministic \emph{guarded choice} among a finite set of action-prefixed processes ($\sum_{i \in I} \alpha_i . P_i$), usually written as $a_1 . P_1 + \ldots + a_n . P_n$. Special cases of the choice construct are the empty process $\emptyset$ which is the sum of zero processes, and the action prefix $\alpha. P$, which is the sum of one process. Notice that choice is commutative by construction. The actions $\alpha$ are: the \emph{internal step} ($\tau$); the act of \emph{receiving} a signal on channel $a$ (the action $a$); \emph{sending} a signal on a channel ($\bar a$). We omit recursion for simplicity; including it does not change the presented results.
Channels are also called \emph{names}. \emph{Free} names $\fn(-)$ are defined by induction, as usual, for processes and labels. The only binding construct is $(\nu a)P$, in which name $a$ is \emph{bound}.
In the following, let $X$ be the set of CCS processes. 

\begin{definition}\label{def:structural-congruence}
\emph{Structural congruence} is the minimal congruence $\equiv \subseteq X \times X$ including $\alpha$-conversion of $a$ in $(\nu a) P$; commutative monoid axioms for the parallel operator with respect to $0$; the equations $(\nu a) (P \parallel Q) \equiv ((\nu a) P) \parallel Q$, $(\nu a) \emptyset \equiv \emptyset$, $(\nu a)(\nu b) P \equiv (\nu b) (\nu a) P$ for all $P$, $Q$, $a\notin \fn(Q)$ and $b$. 
\end{definition}

The labelled transition system for CCS is presented in Figure \ref{fig:CCS-LTS}. The set $L$ of labels is just the set of prefixes $\alpha$. We write $x \tr \alpha y$ as a shorthand for $(\alpha,y) \in g(x)$.
An LTS with labels from $L$ can be represented as a pair $(X,f)$ where $f$ is a function%
\footnote{Here $\Pow(X)$ is the \emph{(co-variant) power set} of $X$. In coalgebras it is typical to assume a cardinality bound on the size of the subsets, as the functor $\Pow$ does not have a final coalgebra. Indeed, in all the systems we define, $\Pow$ can be replaced by a bounded variant, as all our transition sets are finite or countable. Since we do not use the final coalgebra in this paper, the distinction is immaterial here.}
 from $X$ to $\Pow(L \times X)$.  Let $(X,g)$ be the LTS for CCS.
\begin{figure}
	\begin{center}
			{\small 
				$$
			    {\alpha . P + Q \tr{\alpha} P} \,(pre)
				\quad
				\frac{a \notin \fn(\alpha) \quad P \tr{\alpha} P'}	  
				{(\nu a) P \tr{\alpha} (\nu a) P'} \, (res)
				\quad
				\frac{P \tr{\alpha} P'}
				{P \parallel Q \tr{\alpha} P' \parallel Q} \, (par)
				$$ $$
				\frac{P \tr{\bar c} P' \quad Q \tr{c} Q'}
				{P \parallel Q \tr{\tau} P' \parallel Q'} \, (syn)
				\quad
				\frac{P \equiv Q \quad P' \equiv Q' \quad P \tr{\alpha} P'}
				{Q \tr{\alpha} Q'} \, (str)
				$$}
	\end{center}
	\caption{The LTS describing the operational semantics of CCS.}\label{fig:CCS-LTS}
\end{figure}

\begin{definition}\label{def:CCS-bisimilarity} CCS bisimilarity is the greatest symmetric relation $\sim_g \subseteq X \times X$ such that, if $x \sim_g y$ and $x \tr \alpha x'$, there is $y'$ such that $y \tr \alpha y'$ and $x' \sim_g y'$. 
\end{definition}

Notice that we are defining bisimilarity in one specific system (e.g. CCS), not between states of different systems. This corresponds to a standard categorical notion once recognised that LTSs are $\Pow(L \times -)$-coalgebras.

\begin{definition}
	Given a functor $\T : \Set \to \Set$, a $\T$-coalgebra is a pair $(X,f : X \to \T X)$ where $X$ is a set. A homomorphism of coalgebras from $(X,f)$ to $(Y,g)$ is a function $h : X \to Y$ such that $g \circ h = \T h \circ f$. $\T$-coalgebras and their morphisms form the category $\coalg(\T)$.
\end{definition}

\begin{definition}\label{def:coalgebraic-bisimilarity}
	Given a $\T$-coalgebra $(X,f)$, \emph{coalgebraic bisimilarity} $\sim_f \subseteq X \times X$ is defined by $x \sim_f y \iff \exists (Y,g) . \exists h : (X,f) \to (Y,g) . h(x) = h(y)$. 
\end{definition}

It is well-known that under suitable conditions on $\T$ (that the functor for LTSs respects) bisimilarity as in Definition \ref{def:coalgebraic-bisimilarity} coincides with other coalgebraic notions (see e.g. \cite{Sta11}). We do not discuss the details, but we note that for LTSs and the one for CCS in particular, the relations from Definition \ref{def:CCS-bisimilarity} and \ref{def:coalgebraic-bisimilarity} coincide.

\subsection{A dialgebraic semantics of CCS}
\label{sec:CCS-dialgebra}

In \S \ref{sec:CCS-coalgebra}, we have seen the coalgebraic semantics of CCS.
Now we describe it as an $(\F,\B)$-dialgebra for the functors of Definition \ref{def:int-and-obs-functors}.

\begin{definition}\label{def:CCS-dialgebra}
Let $X$ be the set of CCS processes. The dialgebra $(X,f)$ for CCS is the least function obeying to the rules in Figure \ref{fig:CCS-dialgebra}.
\end{definition}

\begin{figure}
\begin{scriptsize}
 \begin{centering}
     $$     
     {\tau . P + Q \utr P}\,(tau)
     \quad 
     \frac{P \utr P'}
     {(\nu a) P \utr (\nu a) P'} \, (res) 
     \quad
     \frac{P \utr P'}{P \parallel Q \utr P' \parallel Q}\,(par_1) 
     \quad
     \frac{(P,Q) \utr R}
     {P \parallel Q \utr R} \, (int) 
     $$         
  \hdashrule[0.2ex]{.4\textwidth}{.3pt}{3mm}   
     $$
     \frac{(P,Q) \utr R \quad a \notin \fn(Q) }
          {((\nu a) P,Q) \utr (\nu a) R }\,(hid)
     \quad
     \frac{(P,Q) \utr R}
          {(P \parallel S,Q) \utr S \parallel R}\,(par_2)
     \quad
          {(\bar a.P + S,a.Q + T) \utr P \parallel Q}\,(syn) 
     $$         
     \hdashrule[0.2ex]{.4\textwidth}{.3pt}{3mm}   
     $$
     \frac{(P,Q) \utr R}
          {(Q,P) \utr R} \, (sym)
     \quad 
     \frac{P \utr R\quad P \equiv Q, R \equiv S}
          {Q \utr S} \,(str_1)
     \quad
     \frac{(P,Q) \utr R\quad P \equiv S,Q \equiv T,R \equiv U}
          {(S,T)\utr U}\,(str_2)
      $$
  \end{centering}
\end{scriptsize}
\caption{The dialgebra for CCS.}\label{fig:CCS-dialgebra}
\end{figure}
	
We briefly comment on the rules. The first group deals with processes in isolation: rule $(tau)$ permits internal computation actions to be executed; Rule $(res)$ allows a process in the scope of a restriction to progress; Rule $(par_1)$ allows one component in a parallel composition to progress independently from the others;  Rule $(int)$ allows any possible interaction between two processes to also happen between internal components of a process in isolation.

The second group of rules defines the semantics of interaction. Rule $(hid)$ permits interaction between a process $P$ in the scope of a restriction, and any other process $Q$, provided that $a$ is not known by $Q$. Recall that the restricted name $a$ can always be $\alpha$-converted to one which is fresh in $Q$. Rule $(par_2)$ allows parallel components of a process $P$ to interact with $Q$ independently from each other. Rule $(syn)$ implements synchronisation between two processes.

Rules $(sym)$, $(str_1)$, $(str_2)$ simplify the definition; alternatively, one can add variants of the other rules taking into account the effects of these three schemes.


\subsection{Comparing the coalgebraic and dialgebraic semantics of CCS}
\label{sec:ccs-comparison}

In this section, we use Theorem \ref{thm:specific-comparisons} to compare the semantics for CCS from Definition \ref{def:CCS-dialgebra} to bisimilarity in the well-known labelled transition system. 

First, we attempt to give some intuition on the constructions of \S\ref{sec:comparing-all}. Consider defining a coalgebra $(X,f')$ for the functor of \S \ref{sec:CCS-coalgebra} out of the dialgebra for CCS of \S \ref{sec:CCS-dialgebra}. We can employ ``witness processes'' such as $a.0$ in experiments such as $(x,a.0)$. For each  $x' \in f(x,a.0)$, we let $f'(x)$ contain the labelled transition $(\bar a,x')$. The idea sounds promising, but in the process we need to refer to specific elements of $X$, namely the witness processes, thus to the specific set $X$ of elements. Natural transformations are not allowed to depend upon an object in this way; therefore, we need the the theory of \S \ref{sec:comparing-all}.

The dialgebra of \S \ref{sec:CCS-dialgebra} describes processes as they interact, with no explicit notion of side effect. The coalgebra of \S \ref{sec:CCS-coalgebra} describes processes in isolation, and their side effects.
It is not difficult to imagine how the two kinds of semantics can be compared. In one direction, starting from the coalgebra, we may define a $(\F,\B)$-dialgebra on the same carrier; for processes in isolation, we run one step of the LTS, and then turn all the $\tau$ transitions into observed results; for interaction between pairs of elements, we run one step of the LTS on each element, and let interaction happen whenever an input (or an output) is matched by the complementing action. In the other direction, starting from the dialgebra, we may define a $\T$-coalgebra, by letting $\tau$ transitions correspond to the observations that are made on a process in isolation, and by running experiments in which we let a process and a ``witness process'' such as $a.\emptyset$ interact. The resulting transitions are labelled with a corresponding action, e.g., $\bar a$ in our case. 

In the rest of the section we closely implement the above plan. From now on, we let $(X,f)$, $(X,g)$, and the functors $\F,\B$ be the dialgebra and the coalgebra for CCS, and the functors from \S \ref{sec:interaction-and-observation-functors}; we let $\T(X) = \Pow(L \times X)$. Below, we define three invariants $\lambda$, $\mu$, $\delta$, that we shall use to instantiate Theorem~\ref{thm:specific-comparisons} twice. In one direction, we will define the coalgebra $\mu \circ \T f \circ \lambda$ out of the dialgebra $f$. In the other direction, we will get the dialgebra $\delta \circ \F g \circ \id_{\F X}$ from $g$.

\begin{definition}\label{def:CCS-invariants}
	We define $\delta : \F \T X \to \B X$, $\lambda : X \to \T \F X$, $\mu : \T \B X \to \T X$ as:
	$$
	\begin{array}{rcl}
		\delta(e)  & = & \left\{ 
		    \begin{array}{lcl} 
		      \{ x \mid (\tau,x) \in p \}  & \quad & \text{if } e = p \in \T X  \\ 
		      \{ x \parallel y \mid \exists a \in \chan . 
		      			\left ((a,x) \in p_1 \land (\bar a, y) \in p_2 \right ) \lor & & \text{if } e = (p_1,p_2) \in \T X \times \T X \\
		     \qquad \qquad \lor \left ( (\bar a,x) \in p_1 \land (a, y) \in p_2 \right ) \}
		      \end{array} \right. \\
	    \lambda(x) & = & \{ (\tau, x) \} \cup \{ (a,(x,\bar a . \emptyset)) \mid a \in \chan  \} 
	    			\cup \{ (\bar a,(x,a . \emptyset)) \mid a \in \chan \} \\
	    \mu (q) & = & \{ (l,x) \mid \exists q' . (l,q') \in q \land x \in q' \}
	\end{array} 
	$$
\end{definition}

Notice how the definition of $\delta$ uses the fact that $X$ is also the carrier of the initial algebra, therefore the parallel composition $x\parallel y$ is defined. An appropriate choice of $\reach_{(X,g)}$ makes $\delta$ an invariant. Also, the definition of $\lambda$ uses specific elements of $X$, such as $\bar a . \emptyset$. On the other hand, $\mu$ is independent of $X$ and extends to a natural transformation from $\T \B$ to $\T$.

\begin{proposition}\label{prop:invariants-CCS}
 Let $E$ be the subcategory of $\dialg(\F,\B)$ of epis whose domain is $(X,f)$, and $E'$ the subcategory of $\dialg(\Id,\T)$ of epis whose domain is $(X,g)$.
 	Let $\reach_{(X,f)}$ be the coslice $(X,f) / E$, and $\reach_{(X,g)}$ be the full subcategory of $(X,g) / E'$ whose objects $h$ commute with the parallel operator, that is, $h(x) = h(x') \land h(y) = h(y') \implies h(x\parallel y) = h(x'\parallel y')$. Then:
\begin{itemize}	
        \item $\reach_{(X,g)}$ obeys to Definition \ref{def:reach};

	\item $\id_{\F X}$ is invariant for $(X,g)$ and $\reach_{(X,g)}$ from $\bar \F$ to $\F \, \bar{\Id} = \bar \F$; 
	
	\item $\delta$ is invariant for $(X,g)$ and $\reach_{(X,g)}$ from $\F \bar \T$ to $\B$; 
	
	\item $\lambda$ is invariant for $(X,f)$ and $\reach_{(X,f)}$ from $\bar \Id$ to $\T \bar \F$; 
	
	\item $\mu$ is invariant for $(X,f)$ and $\reach_{(X,f)}$ from $\T \bar \B$ to $\bar \T$.  
\end{itemize}
\end{proposition}

In Proposition \ref{prop:invariants-CCS}, $\reach_{(X,g)}$ contains only homomorphisms that commute with the parallel operator, strengthening the hypothesis for invariance, which facilitates the proof.
We can now use Theorem \ref{thm:specific-comparisons}, twice. Let $\G = \F$; we obtain the $(\F,\B)$-dialgebra $(X,g^{\id_{\F X},\delta})$ where $f^{\id_{\F X},\delta} = \delta \circ \F g$. Similarly, let $\G = \T$; then we derive the $(\Id,\T)$-dialgebra, that is, $\T$-coalgebra, $(X,f^{\lambda,\mu})$ with $f^{\lambda,\mu} = \mu \circ \T f \circ \lambda$.

\medskip

So far, we have mapped the dialgebra of CCS into a coalgebra, and the coalgebra into an $(\F,\B)$-dialgebra. However, no link is established between $f$ and $g^{\id,\delta}$, or $g$ and  $f^{\lambda,\mu}$. We conclude the paper by proving coincidence of the two semantics. For this, we need the following lemma. 

\begin{lemma}\label{lem:transition-reaction}
	For all channels $a$ and elements $x,y,z$, the following holds:
		$x \tr \tau y \iff x \utr y$; 
		%
		$x \tr{a} y \iff (x,\bar a.\emptyset) \utr y$; 
		%
		$x \tr{\bar a} y \iff (x,a. \emptyset) \utr y$; 
		%
		$(x,y) \utr z \iff \exists b,x',y' . z \equiv x' \parallel y' \land 
		( (x \tr{b} x' \land y \tr{\bar b} y') \lor		  
		(x \tr{\bar b} x' \land y \tr{b} y') )$. 
\end{lemma}

\begin{proposition}\label{pro:equal}
	Let $(X,f)$ and $(Y,g)$ be the dialgebra and the coalgebra for CCS. We have that $f = g^{id_{\F X}, \delta}$ and $g = f^{\lambda,\mu}$. Therefore, bisimilarity in $(X,f)$ and in $(X,g)$ is the same.
\end{proposition}

\subsection{The $\pi$-calculus}


The $\pi$-calculus \cite{mpw92} is a very well known extension of CCS. The calculus takes network mobility into account by the means of fresh name generation and communication. Bisimilarity in the $\pi$-calculus is non-standard, since it requires side conditions on freshness of names that do not permit one to compare labels just syntactically. A coalgebraic semantics is possible by switching from the category $\Set$ to presheaves (see e.g. \cite{ft99}). In this section we provide a dialgebraic semantics to the calculus. Remarkably, the dialgebra we define lives in $\Set$; the difference from the semantics of CCS is just to add data passing in the rule for synchronisation.
We give a very brief summary of the calculus here. The reader may consult e.g., \cite{San01} for further information. The $\pi$-calculus features data passing, and fresh name creation. Channels and data coincide, giving to the calculus its expressive power. The syntax is as follows.
$$ P ::= \sum_{i \in I} \alpha_i . P_i \mid P_1 \parallel P_2 \mid (\nu a) P \qquad \qquad \alpha ::= \tau \mid a(x) \mid \bar ax$$
Again, we do not introduce recursion, as it does not add to the presentation and complicates proofs.  In the syntax, $a,x,y$ range over a countable set of \emph{channel names}.
The prefix $a(x)$ reads $x$ from channel $a$. Therefore, $x$ is bound in $a(x).P$. The prefix $\bar a x$ sends $x$ on channel $a$. The other constructs have the same informal meaning as in the CCS, and share the same syntax.  We adopt the \emph{early} semantics of the calculus. For space reasons, we omit the definition of the corresponding transition system, which is widely available (see e.g. \cite{San01}, Definition 1.3.2). We just mention that the transitions may have four kinds of labels: $\tau$, $\bar a b$, $a b$, $\bar a (x)$, corresponding to silent actions, output of $b$ on channel $a$, input of $b$ on $a$, and \emph{bound output}, where $b$ is a fresh name. Synchronization with a process doing bound output may only take place when $b$ is fresh in the receiving process, which is obtained by $\alpha$-conversion of $b$.

The close syntactic resemblance between CCS and the $\pi$-calculus is reflected in the dialgebraic semantics we propose, as we only need to change one rule to switch from one semantics to the other. The formal definition uses structural congruence, which is the same as Definition \ref{def:structural-congruence},  with the addition of $\alpha$-conversion of variable $x$ for processes under the scope of an input $a(x)$.

\begin{definition}
Let $X$ be the set of $\pi$-calculus terms. Define the $(\F,\B)$-dialgebra $(X,f)$ importing the rules of Definition \ref{def:CCS-dialgebra}, where Rule $(syn)$ is replaced by 
$$
	{(\bar ab.P + S,a(y).Q + T) \utr P \parallel Q[{}^b/_y]}\quad(com)$$
\end{definition}
%
%
%
Rule $(com)$ models data passing in the usual way. Combined with Rule $(hid)$, it implicitly handles \emph{scope extrusion}, which is one of the most difficult bits of the $\pi$-calculus semantics. For example, for $x$ fresh in $Q$, we have $((\nu x) \bar a x . P, a(y).Q)\utr (\nu x)(P \parallel Q[{}^x/_y])$, by first applying $(hid)$ and then $(com)$. Such a simple treatment of scope extrusion is inherited from the reactive system for the $\pi$-calculus that we are mimicking (see \cite{San01}, Definition 1.2.12). The dialgebraic definition adds to it a non-trivial notion of bisimilarity, that we ought to relate to the standard definition. A direct comparison, as in \S \ref{sec:ccs-comparison}, is not obvious, as the coalgebraic semantics lives in a presheaf category, whereas the dialgebraic semantics is defined in $\Set$. However, we are able to reuse well known results for the $\pi$-calculus to obtain a characterisation theorem. Dialgebraic semantics is easier to compare to \emph{strong barbed equivalence}, defined below, than to bisimilarity. We use the \emph{observability predicate} $P\downarrow _\mu$, for $\mu$ in the form $a$ or $\bar a$, that holds whenever $P$ can perform a communication action $a(x)$ or $\bar a x$, respectively. In turn, strong barbed equivalence is defined in terms of \emph{strong barbed bisimilarity}.

\begin{definition}\label{def:pi-strong-barbed}
\emph{Strong barbed bisimilarity} is the largest symmetric relation $\sbb$ such that whenever $P \sbb Q$, $P \downarrow_\mu \implies Q \downarrow_\mu$, and $P \tr{\tau} P' \implies \exists Q' . Q \tr{\tau} Q'$ with $P' \sbb Q'$. 
We say that processes $P$ and $Q$ are \emph{strong barbed equivalent}, written $P \sbe Q$, if for all $R$, $P \parallel R \sbb Q \parallel R$.
\end{definition}

\begin{theorem}\label{thm:pi-calculus-characterization}
 Dialgebraic bisimilarity coincides with strong barbed equivalence.
\end{theorem}

Finally, Theorem 2.2.9 in \cite{San01} proves that strong barbed equivalence coincides with strong early bisimilarity\footnote{Therein, it is shown that the \emph{matching} prefix of the $\pi$-calculus is not required for this result to hold, thus we omit it for simplicity.}. Thus, as a corollary, we get that dialgebraic bisimilarity coincides with early bisimilarity.
%

\noindent

\section{Conclusions and future work}

The most important difference between coalgebras and dialgebras is that there is no final dialgebra, therefore no universal model. This forces one to reason in terms of quotients. The locality which is intrinsic to the definition of a dialgebra deserves in our opinion a more thorough investigation in various directions.

A fundamental problem is to spell out an inductive definition principle, in order to obtain simpler definitions, and compositionality. A conjecture on how to generalise the use of distributive laws \cite{tp97} from bialgebras to dialgebras has been formulated in \cite{Blo12}, and will be developed in future work.  

Logical aspects should also be considered. The interplay between adequate logics for dialgebras, and equational logic on terms, may lead to new insights on algebraic and coalgebraic specifications. 

Another matter is the implementation and verification of dialgebras. Coalgebras have an associated partition refinement procedure that computes the bisimilarity quotient of a system, by the means of iteration along the terminal sequence of the functor $\T$. A generalisation of this procedure to dialgebras appears in \cite{Blo12}, and will be explained and enhanced in future work.

Finally, an open question is the definition of a proof principle for dialgebras, generalising induction and coinduction.

\bibliographystyle{splncs}
\small{\vskip -20pt \bibliography{reactive.bib}}

\begin{thebibliography}{10}

\bibitem{ft99}
Fiore, M.P., Turi, D.:
\newblock Semantics of name and value passing.
\newblock In: 12th Annual Symposium on Logic in Computer Science (LICS), IEEE
  (2001)  93--104

\bibitem{TG69}
Trnkov\'a, V., Goralc\'ik, P.:
\newblock On products in generalized algebraic categories.
\newblock Commentationes Mathematicae Universitatis Carolinae \textbf{010}(1)
  (1969)  49--89

\bibitem{Ada76}
Adámek, J.:
\newblock Limits and colimits in generalized algebraic categories.
\newblock Czechoslovak Mathematical Journal \textbf{26}(1) (1976)  55--64

\bibitem{Hag87}
Hagino, T.:
\newblock A Categorical Programming Language.
\newblock PhD thesis, University of Edinburgh (1987)

\bibitem{PZ01}
Poll, E., Zwanenburg, J.:
\newblock From algebras and coalgebras to dialgebras.
\newblock Electronic Notes in Theoretical Computer Science \textbf{44}(1)
  (2001)  289 -- 307

\bibitem{Rut00}
Rutten, J.J.M.M.:
\newblock Universal coalgebra: a theory of systems.
\newblock Theoretical Computer Science \textbf{249}(1) (2000)  3 -- 80

\bibitem{Vou10}
Voutsadakis, G.:
\newblock Universal dialgebra: unifying algebra and coalgebra.
\newblock Far East Journal of Mathematical Sciences \textbf{44}(1) (2010)
  1--53

\bibitem{CiE11}
Madeira, A., Martins, M.A., Barbosa, L.:
\newblock Models as arrows: the role of dialgebras.
\newblock 7th Conference on Computability in Europe, Sofia, Bulgaria (2011)
  144--153

\bibitem{Cia11}
Ciancia, V.:
\newblock Interaction and observation, categorically.
\newblock In: 4th Interaction and Concurrency Experience. Volume~59 of EPTCS.
  (2011)  25--36

\bibitem{lm00}
Leifer, J.J., Milner, R.:
\newblock Deriving bisimulation congruences for reactive systems.
\newblock In: 11th International Conference on Concurrency Theory (CONCUR).
  Volume 1877 of LNCS., Springer (2000)  243--258

\bibitem{Sew02}
Sewell, P.:
\newblock From rewrite rules to bisimulation congruences.
\newblock Theoretical Computer Science \textbf{274} (March 2002)  183--230

\bibitem{ss03}
Sassone, V., Sobocinski, P.:
\newblock Deriving bisimulation congruences using 2-categories.
\newblock Nordic Journal of Computing \textbf{10}(2) (2003)  163--186

\bibitem{Mil82}
Milner, R.:
\newblock A Calculus of Communicating Systems.
\newblock Springer (1982)

\bibitem{Sta11}
Staton, S.:
\newblock Relating coalgebraic notions of bisimulation.
\newblock Logical Methods in Computer Science \textbf{7}(1) (2011)

\bibitem{mpw92}
Milner, R., Parrow, J., Walker, D.:
\newblock {A Calculus of Mobile Processes, Part I}.
\newblock Information and Computation \textbf{100}(1) (1992)  1--40

\bibitem{San01}
Sangiorgi, D., Walker, D.:
\newblock The {$\pi$}-Calculus: A Theory of Mobile Processes.
\newblock Cambridge University Press, New York, NY, USA (2001)

\bibitem{tp97}
Turi, D., Plotkin, G.:
\newblock Towards a mathematical operational semantics.
\newblock In: 12th Annual Symposium on Logic in Computer Science (LICS), IEEE
  (1997)  280--291

\bibitem{Blo12}
Blok, A.:
\newblock Interaction, observation and denotation: A study of dialgebras for
  program semantics.
\newblock Master's thesis, University of Amsterdam (2012)

\end{thebibliography}

\appendix

\section{Proofs}

\begin{proof}(Proposition \ref{prop:epi-mono-factorisations})
	Consider a dialgebra homomorphism $h  : (X,f) \to (Y,g)$ and the diagram of Figure \ref{fig:dialgebra-homomorphism}. First notice that the epimorphisms in $\dialg(\F,\B)$ are just the epimorphisms in $\Set$ that are dialgebra morphisms.
	Since $\Set$ has epi-mono factorisations, we can factor every arrow $j$ in the diagram, including $h$ itself, as $j_m \circ j_e$ where $j_e$ is epic, and $j_m$ is monic. Call $X'$ the codomain of $h_e$. We are going to endow $X'$ with a dialgebra structure $(X',f')$ so that $h_e$ is a dialgebra morphism from $(X,f)$ to $(X',f')$. 
	
	Since all set functors preserve all epis, and all monos except possibly those with an empty domain, assume that either $\F$ and $\B$ preserve the monos with an empty domain, or that $X \neq \emptyset$. In the latter case, $X' \neq \emptyset$ as there are no arrows whose codomain is the empty set. Thus, $\F h_e$ epic, and $\F h_m$ monic. Then by uniqueness of epi-mono factorisations, $\F h = \F (h_m \circ h_e) = \F h_m \circ \F h_e$ is the epi-mono factorisation of $\F h$. Similarly $\B h = \B h_m \circ \B h_e$ is the unique factorisation of $\B h$. Consider the pushout $P$ of $f_e$, $\F h_e$, and the injection $f'_e : \F X' \to P$. There is an unique commuting arrow $p : P \to \B Y$. Call $p' : \F X \to P$ the diagonal of the pushout. All the arrows we mentioned except $p$ are epi. Similarly, consider the pullback $Q$ of $g_m$, $\B h_m$. There is a unique commuting arrow $q : \F X \to Q$. Also consider the projection $g'_m : Q \to \B X'$ of the pullback, and the diagonal $q' : Q \to \B X$. All the arrows in 
this sub-diagram except $q$ are mono. We have two morphisms $r = p \circ p'$ and $s = q' \circ q$ that commute with the outer square, therefore they are equal. Now factor $p = p_m \circ p_e$ and $q = q_m \circ q_e$. Then $r = p_m \circ p_e \circ p'$ with $p_e \circ p'$ epi. Also $s = q' \circ q_m \circ q_e$ with $q' \circ q_m$ mono. Therefore we have two epi-mono factorisations  of the arrow $r = s$, thus there is an isomorphism $i : \cod(p_e) \to \dom(q_m)$. The composite $f' = g'_m \circ q_m \circ i \circ p_e \circ f'_e$ is an arrow from $\F X'$ to $\G X'$. It is easy to see that $h_e$ is a dialgebra homomorphism from $(X,f)$ to $(X',f')$. Commutativity is by diagram chasing, and uniqueness by $\F h_e$ epi.
\end{proof}

\begin{proof}(Proposition \ref{pro:bisimilarity-equivalence})
    If $k(x) = k(y)$ then $x$ and $y$ are bisimilar by definition. Conversely, given a homomorphism $h$ such that $h(x) = h(y)$, by epi-mono factorisations we can assume $h$ is epic ($X$ is certainly non-empty since we assume an element $x$); w.l.o.g. assume $h$ is a quotient (as bisimilarity is the same in a class of isomorphic epis). Thus, there is an epi $h'$ such that $h' \circ h = k$, thus $k(x) = k(y)$. As bisimilarity is the kernel of a morphism, it is an equivalence.
\end{proof}

\begin{proof}(Proposition \ref{pro:existence-bisimilarity-quotients})
  Call $P$ the cocone of epimorphisms from $(X,f)$. First, observe that $P$ is a small diagram, that is, its objects and morphisms form sets. This is since the cone of of $X$ in $\Set$ forms a set $S$, and $P$ can be at most as large as the product $S \times \B X^{\F X}$ (the possible quotients are as many as the dialgebras whose carrier is a quotient of $X$ in $\Set$). Since $\Set$ is cocomplete, the pushout $Q$ of the morphisms in $P$ exists in $\Set$.  It is not difficult to see that the category of $(\F,\B)$-dialgebras has all colimits that the base category ($\Set$ in our case) has and $\F$ preserves (see Theorem 14 in \cite{Vou10}). Therefore if $\F$ preserves wide pushouts, there is a pushout $(Q,q)$ of $P$ as a diagram in $\dialg(\F,\B)$. The carrier is $Q$; the dialgebra map $q$ is the unique commuting morphism from $\F Q$, which is a pushout of all the $\F h$ for $h$ in $P$, to $\B Q$, that forms a commuting diagram with all the dialgebras that are codomains of morphisms in $P$.
\end{proof}

\begin{proof}(Proposition \ref{pro:int-obs-epi-mono-bisim-quotient})
	$\F$ preserves the initial object and $\B$ sends it into $1$. Therefore they both preserve the specified monos.  By relying on the fact that all epi split, it is easy to see that $\F$ preserves binary pushouts of epis, even though it fails to preserve pushouts in general. Since $\F$ also preserves filtered colimits, it preserves wide pushouts of epis. 
\end{proof}


\begin{proof}(Theorem \ref{thm:bisimilarity-back-and-forth})
 For one direction of the proof, let $\R$ be the kernel of $h : (X,f) \to (Y,g)$. First, let $(x_1,x_2) \in \R$, and consider the case $x_1 \utr z_1$. By commutativity we have $\B h \circ f = g \circ \F h$, thus by definition of $\F$, $\Pow(h)(f(x_1)) = g(h(x_1)) = g(h(x_2)) = \Pow(h)(f(x_2))$, that is $\{ h(z_1) | x_1 \utr z_1 \} = \{ h(z_2) | x_2 \utr z_2 \}$, from which, for the specific $z_1$ that we introduced, there is a $z_2$ such that $x_2 \utr z_2$ and $h(z_1) = h(z_2)$, that is, $(z_1,z_2) \in \R$. Similarly, let $(x_1,x_2),(y_1,y_2) \in \R$, with $(x_1,y_1) \utr z_1$. By commutativity and by definition of $\F$ we have $\Pow(h)(f(x_1,y_1)) = g(h(x_1),h(y_1)) = g(h(x_2),h(y_2)) = \Pow(h)(f(x_2,y_2))$. Therefore, there is $z_2$ such that $(x_2,y_2) \utr z_2$ with $(z_1,z_2) \in \R$.
 
 For the other direction of the proof, assume a dialgebra $(X,f)$, and an equivalence relation $\R$ as in the thesis. We shall define a dialgebra $(Y,g)$ and homomorphism $h : (X,f) \to (Y,g)$ such that $(x_1,x_2) \in \R \iff h(x_1) = h(x_2)$. Let $Y = X _{/_\R}$ (the quotient of $X$ by $\R$). Let $[x]$ denote the equivalence class of $x$, and define $h(x) = [x]$, therefore identifying exactly the processes that are equivalent by $\R$. To conclude the proof, we need to define a commuting dialgebra. Let $g : \F Y \to \B Y$ be defined by $g([x]) = \{ [z] | (x,x') \in \R \land z \in f(x') \}$, $g([x],[y]) = \{ [z] | (x,x') \in \R \land (y,y')\in \R \land z \in f(x',y')\}$. The commutativity requirement on $h$ is now reduced to prove that $g([x]) = \{ [z] | z \in f(x)\}$ and $g([x],[y]) = \{ [z] | z \in f(x,y)\}$. This follows immediately by the hypotesis on $\R$.
\end{proof}

\begin{proof}(Theorem \ref{thm:general-comparisons})
	For all dialgebra morphisms $h$, the diagram in Fig \ref{fig:natural-in-set} commutes. The middle square commutes since $h$ is a dialgebra homomorphism and $\G$ is a functor. The left and right ones by naturality of $\lambda$ and $\mu$. Therefore $h$ is a dialgebra homomorphism from $\K(X,f)$ to $\K(Y,g)$. $K$ is concrete by definition and easily checked to be a functor.
\end{proof}

\begin{proof}(Theorem \ref{pro:nat-uniquely-determined})
	Consider any object $h : (X,f) \to (Y,g)$ of $\reach_{(X,f)}$. We show that $\delta_h$ is uniquely determined by $\delta_{id_{(X,f)}}$. 
	Consider the commuting square
	\begin{center}
		\begin{tikzpicture}
			\obj{FX}{}{\F (\id_{(X,f)})}
			\obj{FY}{below of = FX}{\F h}
			\obj{DUMMY}{right of = FX}{}
			\obj{GX}{right of = DUMMY}{\G (\id_{(X,f)})}
			\obj{GY}{below of = GX}{\G h}
			\arr{\F \hat h}{FX}{FY}
			\arr{\G \hat h}{GX}{GY}
			\arr{\delta_{\id_{(X,f)}}}{FX}{GX}
\			\arr{\delta_h}{FY}{GY}
	\end{tikzpicture}	
	\end{center} 
	Since $\hat h$ is epic, and $\F$ preserves epis, a commuting $\delta_{h}$ is uniquely determined by $\delta_{\id_{(X,f)}}$: by commutativity, for all $x \in \F (\id_{(X,f)})$, we have $\delta_{h}(\F \hat h(x)) = \G \hat h (\delta_{\id_{(X,f)}}(x))$.  If $\id_{(X,f)}$ is invariant, this equation defines a function $\delta_h$. Conversely, if $\delta$ is natural, then $\delta_{\id_{(X,f)}}$ is invariant by definition.
\end{proof}

\begin{proof}(Theorem \ref{thm:specific-comparisons})
Since $\lambda$ and $\mu$ are invariant, they determine corresponding natural transformations by Proposition \ref{pro:nat-uniquely-determined}. If two elements $x_1$ and $x_2$ are bisimilar in $(X,f)$, then there is a morphism of $(\F,\B)$-dialgebras $h : (X,f) \to (Y,g)$ such that $h(x_1) = h(x_2)$. By epi-mono factorisations (Proposition \ref{prop:epi-mono-factorisations}), we can assume w.l.o.g. that $h$ is epic, and by the condition in Definition \ref{def:reach} we can also assume that it is an object of $\reach_{(X,f)}$ (formally one sees that there is an object in the subcategory that identifies $x_1$ and $x_2$). Let $g' = \mu_h \circ \G g \circ \lambda_h$. Notice that $(Y,g')$ is an $(\F',\B')$-dialgebra. Consider the diagram in Figure \ref{fig:natural-in-reach-X-f} and the equations therein. These equations hold by Definition \ref{def:lifting}.  The diagram commutes: the middle square since $h$ is a dialgebra homomorphism and $\G$ a functor, the left and right ones by naturality of $\lambda$ and $\
mu$. From commutativity of the perimeter, we see that $h$ is also a morphism of $(\F',\B')$-dialgebras from $(X,f^{\lambda,\mu})$ to $(Y,g')$, therefore $x_1$ and $x_2$ are bisimilar in $(X,f^{\lambda,\mu})$.
\end{proof}

\begin{proof}(Proposition \ref{prop:invariants-CCS})
	Note that $\reach_{(X,g)}$ respects the conditions in Definition \ref{def:reach} because both the required identity and the final coalgebra (that is, the final $(\Id, \T)$-dialgebra) are included in it. The result on the identity function being invariant is obvious. 
	
	For $\delta$, suppose $\F \T h (e_1) = \F \T h (e_2)$. By definition of $\F$, this implies either $e_1 = p, e_2 = p'$ or $e_1 = (p_1,p_2), e_2 = (p'_1,p'_2)$, where $p,p',p_1,p_2,p'_1,p'_2 \in \T X$.
	
	When $e_1 = p, e_2 = p'$, we have $\T h (p) = \T h (p')$. Thus $\B h (\delta(p)) = \{ h(x) \mid (\tau,x) \in p \} = \{ x | (\tau, x) \in \T h (p) \} = \{ x | (\tau, x) \in \T h (p') \} = \B h (\delta(p'))$. 
	
	In the other case, we have $\B h (\delta (p_1,p_2)) = \{ h(x \parallel y ) \mid \Gamma(p_1,p_2,x,y) \}$ where $\Gamma$ is a shorthand for the defining condition of the set. Consider the set of pairs $S = \{ (h(x),h(y)) \mid \Gamma (p_1,p_2,x,y) \}$.  Under the hypothesis $\F \T h (e_1) = \F \T h (e_2)$, we have $S = \{ (x,y) \mid \Gamma (\T h (p_1),\T h (p_2), x, y) \} = \{ h(x), h(y) \mid \Gamma(p'_1,p'_2,x,y) \}$. We shall prove that if this equality holds, then $\B h (\delta(p_1,p_2)) = \B h (\delta(p'_1,p'_2))$. The equality implies that, for all $x,y$ such that $\Gamma(p_1,p_2,x,y)$, there are $x',y'$ such that $\Gamma(p_1,p_2,x',y')$, and $h(x)=h(x'),h(y)=h(y')$. Therefore, by how we chose the objects of $\reach_{(X,g)}$, we have $h(x\parallel y) = h(x' \parallel y')$. 
	
	The cases for $\lambda$ and $\mu$ are handled similarly.
\end{proof}

\begin{proof}(Lemma \ref{lem:transition-reaction}) The proof is a simple induction on the derivation. Notice that the rules dealing with structural congruence permit us to prove equality of the destination states; otherwise this lemma would hold only up-to structural congruence. We only show the most complicated part of the proof, which is the right to left direction of the last case. Suppose $x \tr{b} x'$ and $y \tr {\bar b} y'$, so we have finite derivations for these transitions. We construct a finite derivation of $(x,y) \utr z$ by induction on the sum of the lengths of the two derivations. The base case is when the last rule of both derivations is $(pre)$; then we can apply Rule $(syn)$ from Figure \ref{fig:CCS-dialgebra} and obtain the thesis. Otherwise, we look at the last step in the derivation of $x \tr b x'$. If either $(res)$, $(par)$ or $(str)$ from Figure \ref{fig:CCS-LTS} is used, we apply the inductive hypothesis to the transitions in the premises, obtaining a derivation of a transition in the 
dialgebra, and then we conclude the derivation of $(x,y) \utr z$ either by Rule $(hid)$, $(par_2)$ or $(str_2)$ from Figure \ref{fig:CCS-dialgebra}, respectively. For $(hid)$ to be applied correctly, we may assume that all bound names are sufficiently fresh, because of $\alpha$-conversion. When Rule $(pre)$ is the last rule used to derive $x \tr b x'$, we construct the derivation of $(x,y) \utr z$ by applying Rule $(sym)$, and then by applying the inductive hypothesis (notice that the argument is symmetric in the polarity of a label (input or output).
\end{proof}

\begin{proof}(Proposition \ref{pro:equal})
	Let us look at equality of $(X,g)$ and $(X,f^{\lambda,\mu})$.  We have $f(\lambda(x)) = \{ (\tau, f(x)) \} \cup \{ (a,f(x,\bar a . \emptyset)) \mid a \in \chan \} \cup \{(\bar a,f(x,a . \emptyset)) \mid a \in \chan \}$. Then, by Lemma \ref{lem:transition-reaction}, this set is equal to $\{ x' \mid x \tr{\tau} x' \} \cup \{(a,\{ x' \mid x \tr {a} x' \}) \mid a \in \chan \}\cup \{(\bar a,\{ x' \mid x \tr {\bar a} x' \} \mid c \in \chan \} $. Then it is immediate that $f^{\lambda,\mu}(x) = \mu(\T f(\lambda(x))) = g(x)$.
	For equality of $(X,f)$ and $(X,g^{\id,\delta})$, we have $\delta(\F g(x)) = \{ x \mid x \tr \tau x' \}$, which by Lemma \ref{lem:transition-reaction} is equal to $\{ x \mid x \utr x'\} = f(x)$. The binary case is analogous, using the last case of Lemma \ref{lem:transition-reaction}.
\end{proof}

\begin{proof}(Theorem \ref{thm:pi-calculus-characterization})
We need to establish some facts before the main proof. 

\medskip
First, notice that using Theorem 2.2.9 in \cite{San01} and well-known properties of the $\pi$-calculus (in particular, compositionality under parallel contexts), Definition \ref{def:pi-strong-barbed} can be given a coinductive flavour. The formal statement is as follows. Strong barbed equivalence is the greatest symmetric relation $\R$ that falls under the hypothesis:
  whenever $P \R Q$, we have $P \downarrow_\mu \implies Q \downarrow_\mu$, and for all $R_1 \R R_2$, we have $P \parallel R_1 \tr{\tau} P' \implies \exists Q' . Q \parallel R_2 \tr{\tau} Q' \land P' \R Q'$.
 To see that $\sbe$ falls under this condition, let $\stackrel{e}{\sim}$ be strong bisimilarity in the $\pi$-calculus, and observe that  $P \sbe Q \implies P \stackrel{e}{\sim} Q \implies \forall R_1 \stackrel{e}{\sim} R_2 . P \parallel R_1 \stackrel{e}{\sim} Q \parallel R_2 \implies \forall R_1 \stackrel{e}{\sim} R_2 . (P \parallel R_1 \tr\tau P' \implies \exists Q' . Q \parallel R_2 \tr \tau Q' \land P' \stackrel{e}{\sim} Q') \implies  \forall R_1 \sbe R_2 . (P \parallel R_1 \tr \tau P' \implies \exists Q' . Q \parallel R_2 \tr \tau Q' \land P' \sbe Q')$. For the other direction, given $P \R Q$, for all $R$, we show that $P\parallel R \sbb Q \parallel R$, by definition of $\R$, with $R_1 = R_2 = R$.

 \medskip 
 \noindent 
 Furthermore, we use the fact that dialgebraic bisimilarity for the $\pi$-calculus is a congruence with respect to parallel composition: $P \sim Q \implies \forall R . P \parallel R \sim Q \parallel R$. This is proved by induction on the rules, using Theorem \ref{thm:bisimilarity-back-and-forth}; the most important case of the proof is to see that, when $P \sim Q$, $(P \parallel R, S) \utr T \implies (Q \parallel R, S) \utr U$ and $T \sim U$. This is immediate from Rule $(par_2)$ in the semantics.
 
\medskip 
 \noindent Finally, the proof uses 
 %
 a variant of the \emph{harmony lemma} 
 (\cite{San01}, Lemma 1.4.15); our statement is $P \tr \tau R \iff P \utr R$ where $P \utr R$ is our dialgebraic semantics. The proof of this fact reuses the standard harmony lemma as follows. First assume Lemma 1.4.15 in \cite{San01}, asserting that the reductions in the $\pi$-calculus (\cite{San01}, Table 1.3) coincide with $\tau$ transitions. Then show that $P \utr R$ can be proved in the dialgebraic semantics if and only if the same reduction can be proved in the standard $\pi$-calculus reduction system. This is done by a straightforward case analysis.

 \medskip 

 \noindent After these preliminary steps, we need to show that $\sbe$ is the same as the bisimilarity quotient $\sim_f$ (in the following, just $\sim$) of the dialgebra $(X,f)$ for the $\pi$-calculus. We proceed by proving first that $\sim$ is included in $\sbe$, then the reverse inclusion.
 
 \noindent First, let us see that $\sim$ is included in $\sbe$ by using our characterization of $\sbe$. Let us assume $P\sim Q$. First we look at the ``barbs''. Let $P \downarrow_{\bar a}$, thus $P \tr{\bar a x} P'$ for some $x$ and $P'$. 
 Then 
 $(P,a(x).\emptyset) \utr P'$. By Theorem \ref{thm:bisimilarity-back-and-forth}, $(Q,a(x).\emptyset) \utr Q'$ with $P' \sim Q'$. Then $Q \downarrow_{\bar a}$. Similarly for the case $P \downarrow_{a}$. 
 
 \medskip

 \noindent Next, consider an arbitrary process $R$, with $P \parallel R \tr \tau P'$. We show $Q \parallel R \sim Q'$ with $P' \sim Q'$. One of three possibilities holds: (1) $R \tr \tau R''$ with $P' = P \parallel R''$; (2) $P \tr \tau P''$ with $P' = P'' \parallel R$; (3) $P$ and $R$ synchronize. In Case 1, we have $Q \parallel R \tr \tau Q \parallel R''$, thus we use compositionality of $\sim$ with respect to parallel composition. In Case 2, by the harmony lemma, we have $P \utr P''$, and by Theorem \ref{thm:bisimilarity-back-and-forth}, $Q \utr Q''$ with $P'' \sim Q''$. Then by compositionality with respect to parallel composition we get the thesis. In Case 3, we have 
 $(P,R) \utr P'$. By Theorem \ref{thm:bisimilarity-back-and-forth} we have $(Q,R) \utr Q'$ with $P' \sim Q'$. Applying Rule $(int)$ in the definition of the dialgebra, $Q \parallel R \utr Q'$. By the harmony lemma $Q \parallel R \tr \tau Q'$, proving the thesis.
 
 \medskip
 
 \noindent For the second half of the proof, consider $P_1 \sbe P_2$, with  $P_1 \utr R_1$. Applying the harmony lemma, $P_1 \tr{\tau} R_1$, thus there is $R_2$ such that $P_2 \tr{\tau} R_2$ and $R_1 \sbe R_2$. The thesis follows by the harmony lemma.
 
 \medskip
 \noindent Finally, let $P_1 \sbe P_2$, $Q_1 \sbe Q_2$, with $(P_1,Q_1) \utr R_1$. Without loss of generality, assume $P_1$ is sending data and $Q_1$ is receiving (the other case is symmetric). Let us import notation from \cite{San01}, shortening $(\nu x_1) \ldots (\nu x_n)$ as $(\nu \tilde x)$ where $\tilde x$ is the set of channels $x_1,\ldots,x_n$. By induction on the rules, 
 it is easy to see that $P_1 \equiv (\nu \tilde x) ((\bar a x. P'_1 + M_1) \parallel N_1)$, $Q_1 \equiv (\nu \tilde y) ((a(x).Q'_1 + S_1) \parallel T_1)$, with $R_1 \equiv (\nu \tilde x) (\nu \tilde y)(P'_1 \parallel N_1 \parallel Q'_1 \parallel T_1)$, where, importantly, $a \notin \tilde x \cup \tilde y$, $x \notin \fn(Q_1) \cup \tilde y$, and besides, ``all fresh names are sufficiently fresh'', that is: $\tilde x \cap \tilde y = \emptyset$, $\fn(P_1) \cap \tilde y = \emptyset$, $\fn(Q_1) \cap \tilde x = \emptyset$. Notice that we deliberately chose to use $x$ in $Q_1$; by doing so, under the previous assumptions, we can avoid writing the substitution in $Q'_1$ explicitly. By rule $(int)$ and the harmony lemma, we have $P_1 \parallel Q_1 \tr \tau R_1$, thus by $\sbe$ we have $P_2 \parallel Q_2 \tr \tau R_2$ with $R_1 \sbe R_2$. 
 Furthermore, $P_1 \parallel a(x).0 \tr \tau (\nu \tilde x) (P'_1 \parallel N_1)$, thus $P_2 \parallel a(x).0 \tr \tau U$ with $U \sbe (\nu \tilde x) P'_1 \parallel N_1$, similarly $Q_2 \parallel \bar a x . 0 \tr \tau V$ with $V \sbe (\nu \tilde y)Q'_1 \parallel T_1$, therefore $R_2 \sbe U \parallel V$. 
 %
 
 \newcommand{\ndownarrow}{/\hskip -7pt \downarrow}
 
 We need to see that $(P_2,Q_2) \utr R_2$; we shall do this by proving that $P_2 \equiv (\nu\tilde z)((\bar a x . P'_2 + M_2)\parallel N_2)$ and $Q_2 \equiv (\nu\tilde t) ((a(x).Q'_2+S_2)\parallel T_2)$, with freshness assumptions similar to the ones for $P_1$ and $Q_2$, and that $U \sbe (\nu\tilde z) (P'_2 \parallel N_2)$, $V \sbe (\nu \tilde t)(Q'_2 \parallel T_2)$. 
 
 For this, we can use barbs and specially crafted contexts.  Let $d$, $e$ be names that are fresh in all the entities mentioned so far. Consider $D = a(m) . ( (\bar d d . 0 + \bar m m . 0) \parallel x(n). \bar e e .0 )$. Roughly, the parallel component that reads on channel $x$ is used to detect when the name $m$ received on $a$ is equal to $x$. This can also be done using matching, that we did not include precisely because of this property. This idea comes from \cite{San01} (see the remarks before the proof of Theorem 2.2.9).
 We have $P_1 \parallel D \tr \tau (\nu \tilde x)(P'_1 \parallel N_1) \parallel ((\bar d d . 0 + \bar x x . 0) \parallel x(n) . \bar e e .0 ) \downarrow_{\bar d} \tr \tau (\nu \tilde x)(P'_1 \parallel N_1) \parallel \bar e e .0 \downarrow_{\bar e} \ndownarrow_{\bar d}$ (where $\ndownarrow_\mu$ means that the observable $\mu$ is not present). Thus by $P_1 \sbe P_2$ we have $P_2 \parallel D \tr \tau W \downarrow_{\bar d}$ with $W \sbe U \parallel  ((\bar d d . 0 + \bar k k . 0) \parallel x(n) . \bar e e .0 )$, where $k$ is a name (received on channel $a$) that we want to prove equal to $x$. Indeed, it suffices to observe that in the empty context, by $\sbe$, $W \tr \tau W'\downarrow _ {\bar e} \ndownarrow_{\bar d}$ which is only possible if $k = x$. We also see that $P_2 \parallel a(m).\bar d d . 0 \tr \tau W'' \sbe U \parallel \bar d d .0$. From this, and inspection of the rules of the $\pi$-calculus, one concludes that $P_2$ is in the requested form. Similarly for $Q_2$.
\end{proof}

\end{document}